\documentclass[11pt,a4paper]{article}

\usepackage{graphicx}
\usepackage{amssymb}

\usepackage{times}
\newcommand{\bea}{\begin{eqnarray}} 
\newcommand{\eea}{\end{eqnarray}}

\usepackage{graphicx} 
\usepackage{amssymb}

\begin{document} 
 
\begin{Large} 
\begin{center} 
\textbf{
Comparing bird and human soaring strategies
\footnote{
Cite this paper as: \newline
Akos Z, Nagy M, Vicsek T (2008) Comparing bird and human soaring strategies. 
{\it Proc Natl Acad Sci USA}, {\bf 105}: 4139-4143.
}
}
\end{center} 
\end{Large} 

\medskip

\begin{large} 
\begin{center} 
Zsuzsa \'Akos$^{\dagger}$, 
M\'at\'e Nagy$^{\dagger}$, and
Tam\'as Vicsek$^{\dagger\ddagger\nmid}$ 
\end{center} 
\end{large} 

\bigskip

$^{\dagger}$Department of Biological Physics, E\"otv\"os University, 
 P\'azm\'any P\'eter s\'et\'any 1A, H-1117 Budapest, Hungary
$^{\ddagger}$Statistical and Biological Physics Research Group 
 of the Hungarian Academy of Sciences, P\'azm\'any P\'eter s\'et\'any 1A,
 H-1117 Budapest, Hungary
$^{\nmid}$To whom correspondence should be addressed. 
 E-mail: vicsek@angel.elte.hu.
 
\bigskip

\begin{abstract}
Gliding saves much energy, and to make large distances
using only this form of flight represents a great challenge 
for both birds and people. The solution is to make use 
of the so-called thermals, which are localized, warmer 
regions in the atmosphere moving upwards with a speed 
exceeding the descent rate of bird and plane. Whereas birds 
use this technique mainly for foraging, humans do it as 
a sporting activity. Thermalling involves efficient 
optimization including the skilful localization of thermals, 
trying to guess the most favorable route, estimating the best 
descending rate, etc. In this study, we address the question 
whether there are any analogies between the solutions birds 
and humans find to handle the above task. High-resolution 
track logs were taken from thermalling falcons and paraglider 
pilots to determine the essential parameters of the flight 
patterns. We find that there are relevant common features 
in the ways birds and humans use thermals. In particular, 
falcons seem to reproduce the MacCready formula widely 
used by gliders to calculate the best slope to take before an 
upcoming thermal.

\medskip 
Supplementary materials are available at the webpage dedicated to this work: 

{\tt http://angel.elte.hu/thermalling/}. 
\end{abstract}
\bigskip 

\begin{figure}[htb]
{
\centerline{\includegraphics[width=8.7cm]{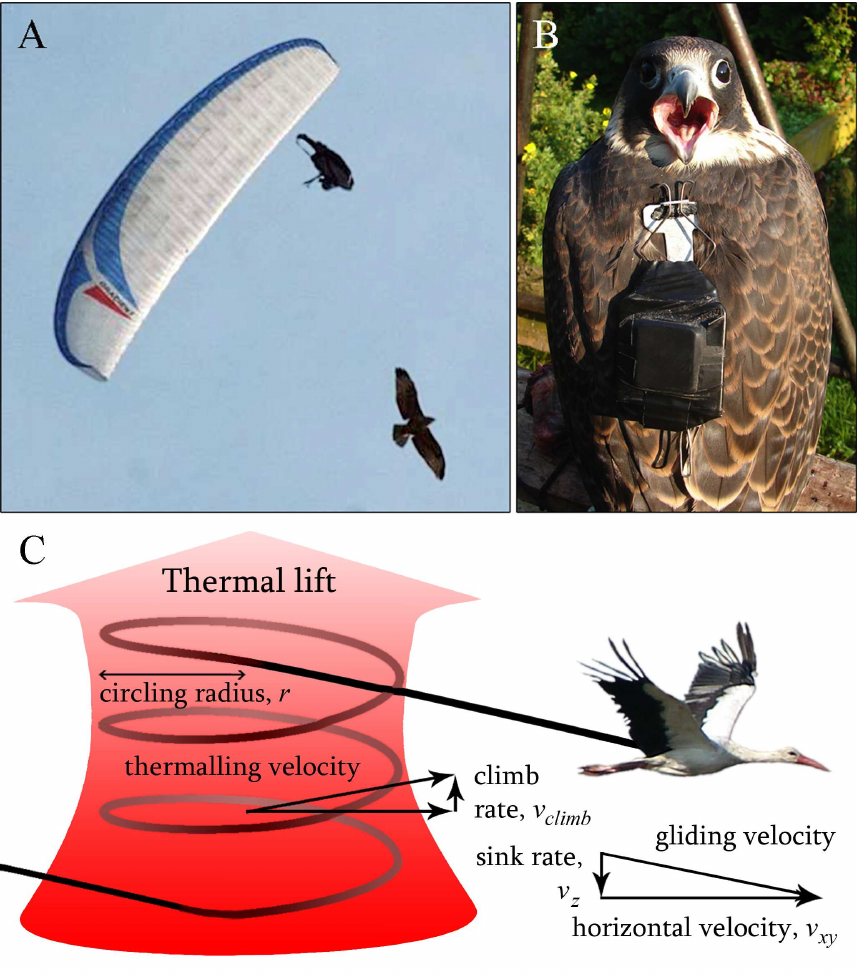}}}
\caption{\baselineskip=0.35truecm{\small
{\bf Photos of the observed flyers and the tracking device.} (A) A paragliding
pilot and a bird of prey thermalling together. (B) Peregrine falcon with
the GPS device on its back. (C) Schematic picture of the thermalling and gliding
parts of the flights with the notations indicated.}}
\label{fig:fig1}
\end{figure}

During long-term gliding, birds and people make use of the
so-called thermals, which are spatially and temporally localized
parts of the atmosphere typically moving upwards with a
speed in the range of $1–5 m/s$. After locating it, a glider remains
within a thermal by circling until the desired height is attained.
Then, a more or less straight advancing, but sinking, phase
follows until the next thermal is reached. Paraglider pilots use
watching the birds thermalling nearby for finding the next
thermal, and sometimes the birds seem to follow the glider (Fig.
\ref{fig:fig1}A). Learning about previously unavailable details of this fascinating
process can lead us to a better understanding of the main
features of flight trajectories and optimization tactics. To locate
the best route to a distant point, at least in the case of human
gliders who typically use specific devices assisting in making the
best decisions, is a complex mental process involving both
calculations and intuition. We consider thermalling as one of the
scarce examples when an intellectually driven activity of humans
is apparently so closely related to the actual behavior of an
animal. Several interesting questions emerge: Does the obvious
size difference result in a different flight pattern and speed? Are
the common tricks the same or are there alternative successful
tactics?

Because collecting data on the soaring flight of birds is a rather
difficult task, several techniques have been used for this purpose. A
powered sailplane with a camera and ornithodolite techniques were
used to determine the polar curves and the circling radius of various
birds \cite{bib1}\cite{bib2}. Gliding of four different bird species was investigated
by radar during their migration \cite{bib3}\cite{bib4}. An altimeter with a satellite
transmitter was used in similar studies on the American White
pelicans in Nevada \cite{bib5}. A further project demonstrated that the
Magnificent frigate bird is thermalling continuously, day and night
\cite{bib6}. Since MacCready published his theory about soaring flight
optimization \cite{bib7}, sailplane pilots have tried to adjust their gliding
speed to the expected thermal climb rate according to their own
polar curve $p(v_{xy})$ (vertical speed versus the horizontal one, $v_{xy}$,
during gliding). In this context, the migration flight of Marsh
harriers was studied \cite{bib8}. Polar curves of several bird species were
measured in wind tunnel studies on trained birds \cite{bib9}. The miniaturization
of GPS devices enabled their usage in bird flight research
(Fig. \ref{fig:fig1}A). A miniaturized GPS device was used to investigate the
navigation strategy of Homing pigeons \cite{bib10}\cite{bib11}. Very recently,
simulations were used to calculate the flight efficiency of the huge
volant bird Argentavis magnificiens from the upper Miocene \cite{bib12},
and wind tunnel experiments and modeling were carried out to
understand how swifts use morphing wings to control their glide
performance \cite{bib13}.

We investigated primarily the soaring flight of the Peregrine
falcon (Falco peregrinus), but we also obtained a smaller set of
data for the White stork (Ciconia ciconia) to see whether there
is a qualitative difference between the two species. Peregrine
falcons use the thermals during foraging to soar up to a suitable
height from where they can stoop for the prey, and they are able
to migrate $\approx 190 km/day$ with this soaring technique \cite{bib14}. White
storks make great use of thermals during their 10,000-km-long
annual migration from the breeding area to the wintering
quarters. We assume that both birds and humans are trying to
maximize their cross-country speed, although for different reasons.
It is beneficial for migrating birds to get to their wintering
(breeding) grounds spending a migration period as short as
possible (during migration, there is a relatively smaller food
supply, and the migrating birds are more vulnerable), whereas in
the case of birds of prey, scanning a larger area for a shorter time
period is more advantageous. The detailed data we have about
humans have been collected from paraglider and hang glider
contests where the pilots are aiming at the highest speed possible
to win the competition. The trajectories were obtained with the
help of an ultra light GPS device providing the three dimensional
position data with a high spatial ($1 m$) and temporal ($1 s$)
resolution (see Materials and Methods), except for the hang
gliders for which positions at every 5 seconds could be downloaded
from flight contest home pages. Thus, smaller scale
details of the flight trajectories could not be obtained for the
hang gliders.

\section{Results and Discussion}

\begin{figure}[htb]
{
\centerline{\includegraphics[width=13.12cm]{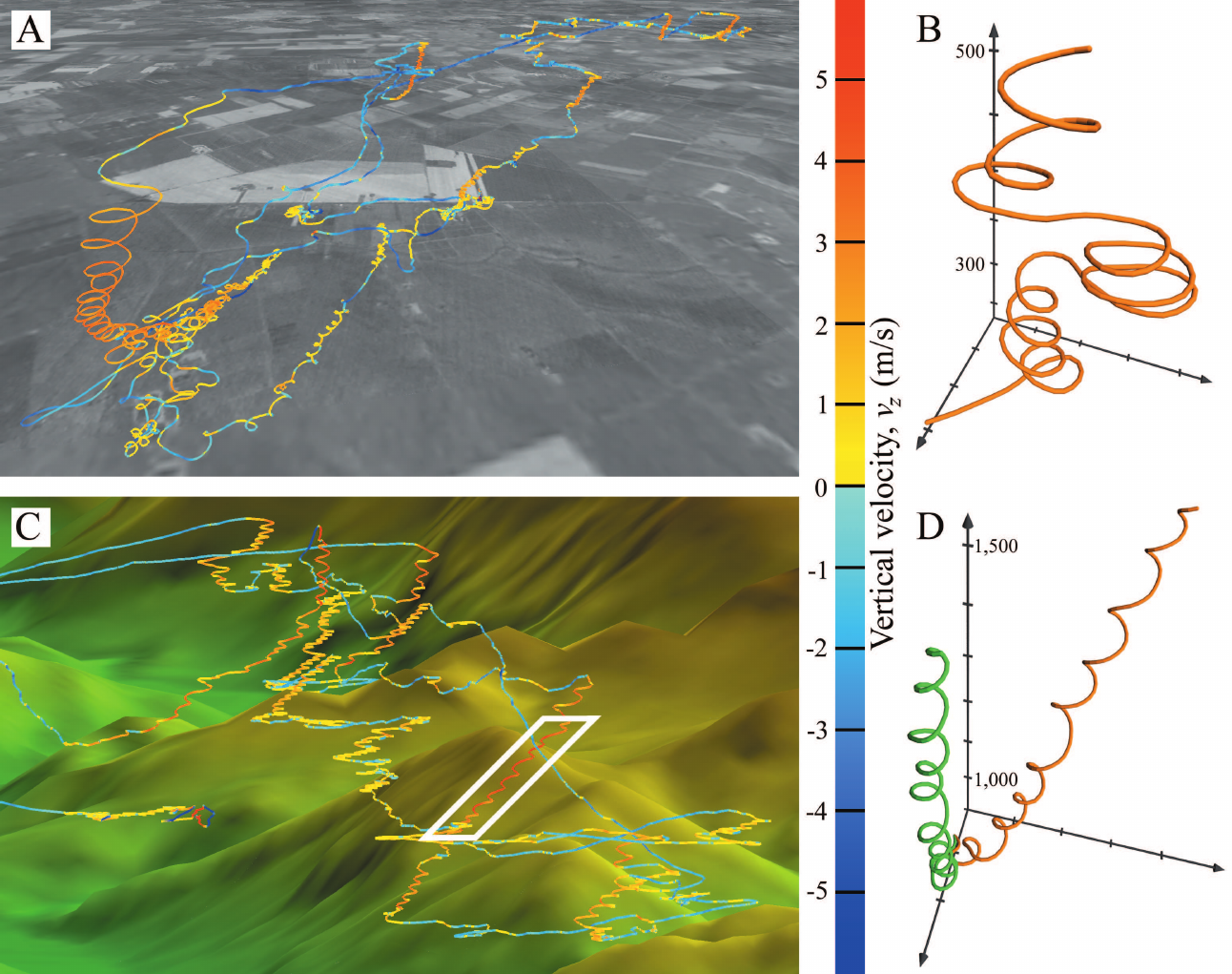}}}
\caption{\baselineskip=0.35truecm{\small
{\bf Visualizations of the track logs of a falcon and a paraglider.} (A) Trajectory 
(track log) of a single flight of the falcon with the background being a black
and white satellite map of the region. Coloring indicates the values of the vertical 
velocity component; red corresponds to climbing (mostly within thermals),
blue to sinking (gliding parts). (B) An interesting feature of the falcon's 
thermalling flight is that from time to time it changes the direction of circling. Marks
on the axes indicate 50-m distances. (C) Trajectory of a paraglider pilot with the local 
relief. (D) Compensating for the wind: the red trajectory corresponds to
the original data (in the white box in C), whereas the same trajectory is shown in 
green after the effect of the wind has been eliminated. Marks on the axes indicate
100-m distances.}}
\label{fig:fig2}
\end{figure}

To characterize the thermalling part of the flights (two examples
are visualized in Fig. \ref{fig:fig2} A and C; see also supporting information
({\it SI}) Movies 1-4), first their drift due to wind had to be
eliminated (Fig. \ref{fig:fig2}D). The horizontal (perpendicular to gravity)
circling radius distribution of these wind-removed trajectories
was calculated. The most frequent circling radius data were given
as the mode value of an Inverse Gaussian function fitted on the
circling radius distribution function, where the data of thermalling
in clockwise ($-$) and counterclockwise ($+$) directions
were calculated separately. Table \ref{tab:table1} shows the results of the
analysis. The circling radius distribution functions indicate that
the typical radius values of the soaring flight of the two bird
species (falcon and stork) and the paragliders are in the range of
20-26 m, {\it e.g.,} do not deviate significantly. The same closeness is
true for the average velocities, indicating that it is rather the
parameters of the representative polar curves (being very similar
in the present case) than other features of the flyers that
influence the flight patterns of birds and paragliders. Our
findings concerning birds are in agreement with the observations
and argument by Pennycuick \cite{bib2}, {\it e.g.,} the mean circling radius is
linearly proportional to wing loading. The same proportionality,
however, does not hold for the paragliders (see {\it SI Appendix} for
details); thus, despite the different values of their wing loadings,
paraglider pilots and the birds considered fly in roughly the same
part of the thermals.

\begin{table}[htb] 
\caption{\baselineskip=0.35truecm{\small 
{\bf Soaring flight data determined for the Peregrine falcon, White stork, and paraglider.}}} 
\centering 
\begin{tabular}{l cccccccc} 
\hline \hline 
Flyer 		& $r_{+}$, m & $r_{-}$, m & $v_{circling}$, m/s & $v_{climb}$, m/s & $v_{xy}$, m/s & $v_{z}$, m/s & $AB$, $^\circ$ &  $T$, s \\
\hline
Falcon 		& $20.2$ & $22.0$ & $10.0$ & $1.4$ & $13.3$ & $-1.9$ & $26$ & $15.4$ \\
Stork 		& $22.3$ & $21.1$ & $9.5$  & $0.9$ & $12.1$ & $-1.1$ & $23$ & $15.5$ \\
Paraglider 	& $24.9$ & $26.0$ & $8.7$  & $1.4$ & $10.9$ & $-1.3$ & $17$ & $20.1$ \\
\hline
\multicolumn{9}{p{15cm}}
{\small $r_{+}$ and $r_{-}$, thermal circling radius in counterclockwise and clockwise directions, respectively; 
$v_{circling}$, average circling velocity, horizontal velocity component during thermalling; 
$v_{climb}$, average climb rate, vertical velocity during thermalling; 
$v_{xy}$, average horizontal velocity during gliding; 
$v_{z}$, average vertical velocity during gliding, sinking velocity, sink rate; 
$AB$, angle of bank; 
$T$, circling period time. }

\bigskip 
\end{tabular} 
\label{tab:table1} 
\end{table}

To check whether the falcon and the paragliders follow the
predictions of the MacCready theory, the knowledge of the
corresponding polar curves is needed. We used polar curves
published by the manufacturer (if it was available) or measured
by pilots. The falcon polar curve was obtained from the data
points measured in wind tunnel by Tucker \cite{bib9}. In addition, for
the falcon, we determined the "effective" polar curve as fitted
to the measured average sinking and horizontal velocities of the
non-thermalling parts of the flights (gray curves). Intermittent
periods of flapping flight (embedded into ordinary gliding) of
the falcon are the likely reason for the larger spread (as
compared with para/hang glider data) of the measured values
around the polar curve. We fitted the polar data by 
$f(x) = a/x + b x^3 + c x^{-3}$, with $a=-4.1$, $b=-0.00056$, and $c=-100$. For
the "effective" polar curve, we used a parabola for simplicity,
$f(x) = a x^2 + b x + c$, with $a=-0.014$, $b=0.2$ and $c=-1.58$.
When determining the polar curve from actual flight data, one
also has to take into account rising and sinking air masses during
the gliding periods, resulting in a higher scattering of the data
(because the falcon was thermalling over flat regions with no
soaring flight over wind-blown ridges, no systematic errors were
expected to occur). The polar curves of the paragliders and hang
gliders were fitted by $f(x) = a x^2 + b x + c$, with $a=-0.015$, $b=0.16$ and 
$c=-1.2$ and $a=-0.0095$, $b=0.2$, $c=-1.7$,
respectively. The resulting root mean squares of the residuals
were in the range of $0.02-0.05$.

\begin{figure}[htb]
{
\centerline{\includegraphics[width=13.51cm]{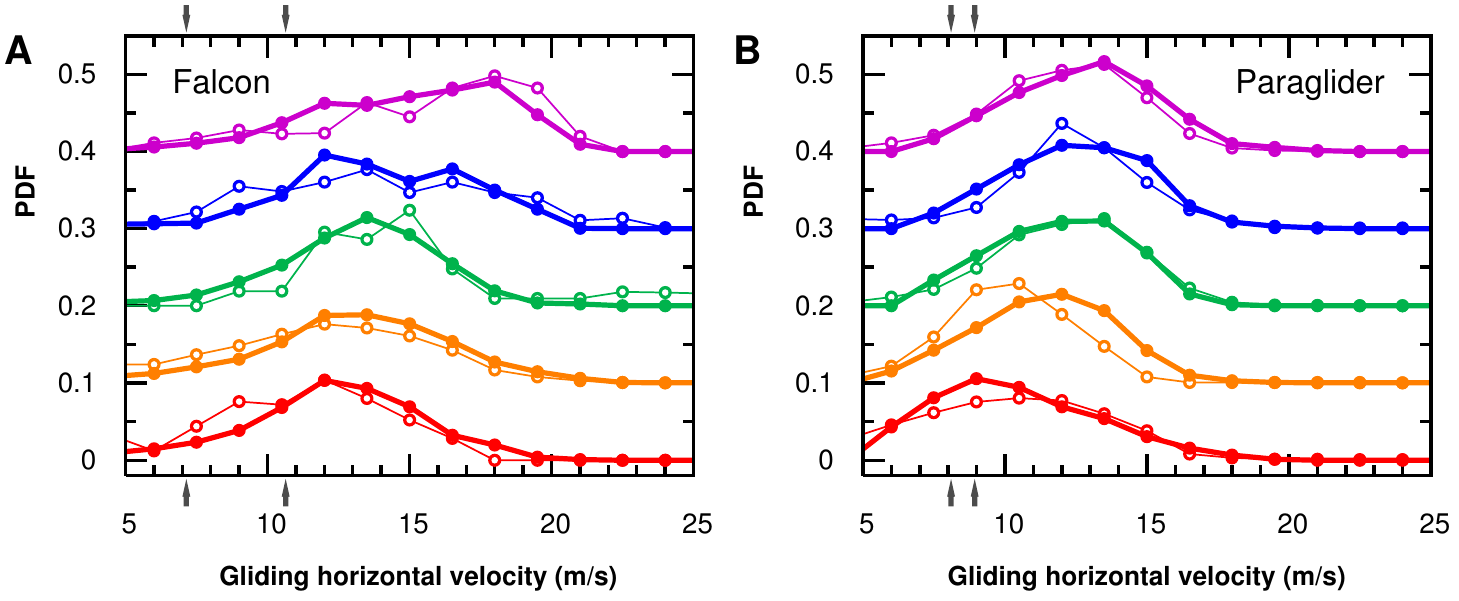}}}
\caption{\baselineskip=0.35truecm{\small
{\bf Comparison of the actual and the predicted (by the MacCready theory) speeds.} 
This figure shows two distributions of the gliding horizontal velocities
during the top 5 days concerning thermalling conditions for the falcon (A) 
and two paragliders (B) who are among the few best performing contestants. 
The data are shifted upward by a value 0.1 for different days to improve the 
visualization. The filled and open symbols denote the circling radius distribution (PDF)
of the predicted and the actual values, respectively. The predicted values were 
obtained by feeding the climbing rate distribution into the MacCready formula
and calculating the corresponding horizontal velocity distribution from it. A 
perfect agreement between the observed velocities and the predicted ones would
correspond to birds and people applying the theory to a 100\%degree. The left 
and the right arrows on the x axis show the horizontal velocity value corresponding
to the minimum sink and the best glide ratio, respectively.}}
\label{fig:fig3}
\end{figure}

Next, we estimated how closely the falcons, the hang glider and
the paraglider pilots follow the optimal thermalling strategy ({\it e.g.,}
the best sinking speed to choose for optimal overall horizontal
velocity) as given by the MacCready theory (in principle there
could be various strategies used by individual flyers but the only
scientifically well documented strategy is this theory and its
variants). Knowing the functional form of the polar curve the
optimal values can be calculated for various climb rates $v_{climb}$
from $v_{climb} = p(v_{xy})-v_{xy} dp(v_{xy})/dv_{xy}$ (see {\it Materials and Methods}).
In Fig. \ref{fig:fig3}, we show a comparison of the actual and the predicted
(by the MacCready theory) velocity distributions for the falcon
and the paragliders. The predicted values were obtained by
feeding the climbing rate distribution into the MacCready formula
and calculating the corresponding horizontal velocity
distribution from it. A perfect agreement between the observed
velocities and the predicted ones would correspond to birds and
people applying the theory to a 100\% degree. We chose to
quantify the agreement by calculating the mean square deviation
of the actual and predicted distributions of the gliding velocities
$v_{xy}$ (where the predicted distributions are obtained by inserting
the measured climbing rates into the MacCready expression and
calculating the corresponding horizontal velocity distribution
from it). Then, this deviation was compared with an average
deviation that would have been observed, if the birds and the
paragliders had used an incorrect value (randomly selected from
other flights) for the climb rate in the preceding thermal when
"calculating" the best theoretical value for the horizontal speed.
The conclusion of this calculation is that, for both the Peregrine
falcon and the paragliders, the actual deviations are significantly
different from those obtained for the randomized case. In
particular, the Student's {\it t} test gives the values $t_{falcon} = 2.36$ and
$t_{paraglider} = 2.80$ for the comparison of the actual and the
randomized data for the falcon's and the paragliders' flights,
respectively. The corresponding t values for a 2.5\% level of
significance are smaller, $t_{2.5, falcon}=2.179$ and $t_{2.5, paraglider}=2.131$
(see {\it SI Appendix}).

\begin{figure}[htb]
{
\centerline{\includegraphics[width=13.5cm]{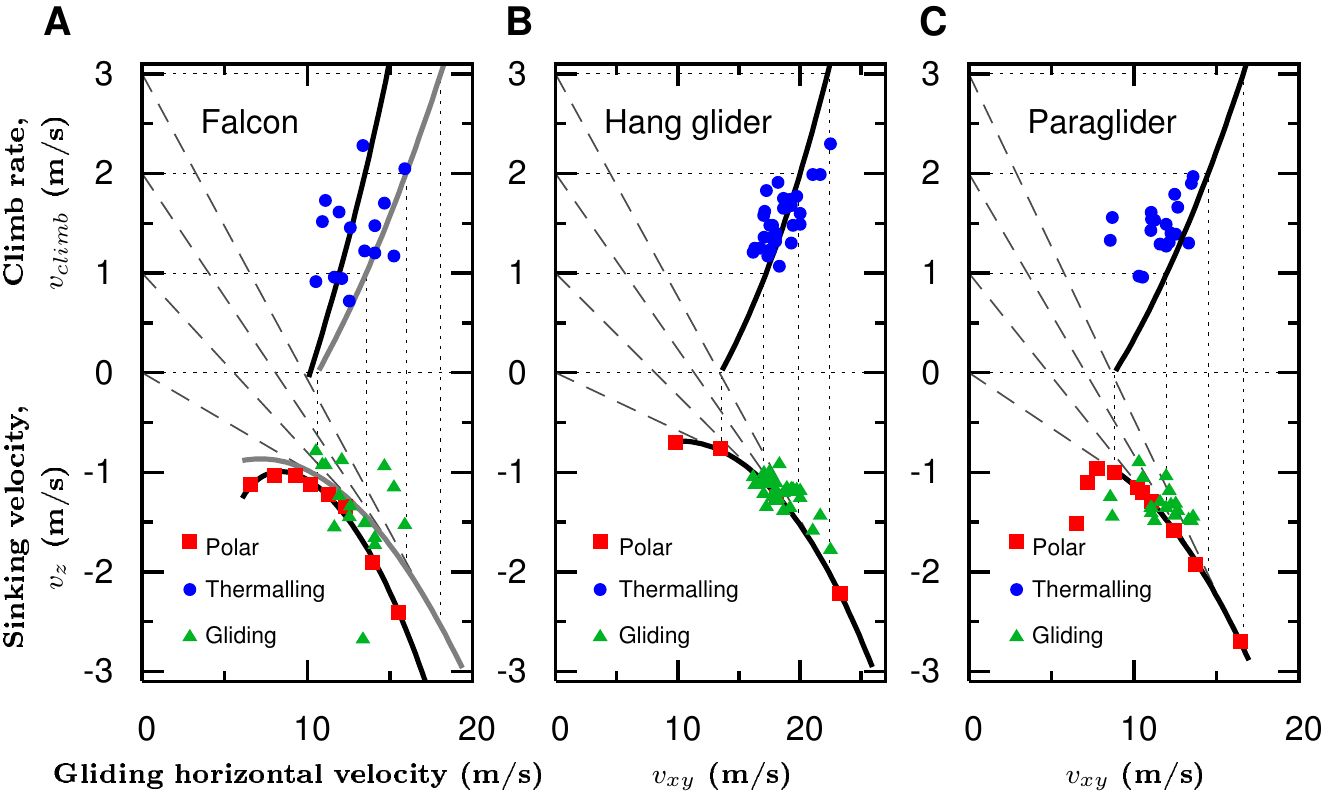}}}
\caption{\baselineskip=0.35truecm{\small
{\bf Soaring strategy plots of flyers.} Squares denote the known points of the polar curve 
(solid line). The solid line in the upper part represents the calculated
optimal soaring strategy curve (see text). The dashed lines show tangent lines to the polar 
curve from points corresponding to different climb rates represented
on the y axis. The dotted lines indicate the horizontal gliding speed values at the tangent 
point of the polar curve for the given dashed lines, and the
corresponding climb rates. Circles denote the measured average thermal climb rate for 
the given measured average gliding horizontal velocity (one circle
represents one flight). Measured average gliding horizontal velocity as a 
function of sinking speed of each gliding flight is indicated by triangles. (A) Peregrine
falcon. The gray curves show the effective polar curve (due to flapping flight 
parts) and the corresponding optimal strategy (see text). (B) The flights of the hang
glider pilot reaching third place. (C) Two paraglider pilots reaching first and second 
place of OLC.}}
\label{fig:fig4}
\end{figure}

Fig. \ref{fig:fig4} shows the polar curve, the calculated optimal soaring
strategy curve and the measured, flight averaged data points for
falcon, a rigid hang glider and for two paraglider pilots (Fig. \ref{fig:fig4}
A, B, and C, respectively) showing a reasonable agreement
between theory and observations. The original MacCready
theory did not take into account several factors that could
influence the optimal choice for the gliding speed. These factors
include changing of the air density as a function of height or the
fact that the various thermals even within a single flight are quite
different as concerning their lifting potential \cite{bib15}. We estimate
that the corresponding accumulated error involved is in the
range of 10-15\%, and conclude that a perturbation of this order
does not change our conclusion. In fact, a range of thermalling
conditions is useful for checking the statistically relevant
applicability of the MacCready theory.

Thus, we found that the leading paraglider pilots taking part in
world contests and the falcon follow a similar flight pattern and a
soaring strategy close to the optimal one as predicted by the theory.
Paraglider pilots apply somewhat slower horizontal gliding speed as
the optimal (points are more scattered to the left of the optimal
curve). This can be interpreted by taking into account that the
paragliders' glide ratio is worse than that of hang gliders, so they
chose a lower speed to minimize the risk of not reaching the next
thermal before landing. In addition, paragliders have a lower
stability at higher speeds, so in some situations, pilots do not apply
the maximum speed for safety reasons. The similarity of the
paragliders' and the falcon's flight strategies could be demonstrated
by considering the effective polar curve we introduced to take into
account the occasional flapping flight periods of the falcon.
As for our original question, the result seems to be a draw. All
of the parameters we determined were nearly the same for both
humans and birds. Thus, as it happens, evolving flight strategies of
birds and human calculations lead to virtually the same outcome.

\section{Materials and Methods}

{\bf The GPS Device.} The GPS device we developed was based on a Fastrax (Vantaa,
Finland) product. It was capable of logging 24,500 log points (latitude, longitude,
and altitude coordinates and time) and had the size of 4.5-6.2cm, weighing only
$34g$. This logging function could be restricted to the time when the bird was
moving. The spatial and temporal resolution of the device was 1m and 1s with
an error of 0.1m and 20ns, respectively. At evenings we detached the GPS device
from the birds and downloaded the log files from the GPS to a PC. With the above
setting, the memory was sufficient for storing the flight data of a whole day.We
constructed four GPS units with the same features. Teflon-treated ribbons from
Marshall Radio Telemetry (North Salt Lake City, UT), commercially used for similar
purposes were used to fix them on the birds.

{\bf Collecting Data About Flights of Tame Birds.} In the Peregrine falcon project,
we cooperated with a professional falconer breeding falcons. His breeding
technique is to let the young birds fly free from the time they first start
flying until they become almost totally self-supporting, which takes $\approx$ 1
month. During this time, the birds come home for food and sleep, and learn
to hunt for themselves. Meanwhile, they become masters of soaring. The
falcon chicks were hand-raised, so they were tame, enabling the falconer
to change the GPS unit on the bird's back. For safety reasons, we also used
a radio device (Micro transmitter by Marshall Radio Telemetry). This project
took place in July 2006, in the Southeast part of Hungary near to the town
of B\'ek\'escsaba. We collected the data for 3 weeks continuously. Fortunately,
the weather was good during this period, with high thermal activity
almost every day.

Under the White stork project, we worked in cooperation with the Hortob\'agy
National Park and one of its institutions, the Bird Repatriation Station. Two
$\approx$ 2-week-old storks that had fallen out of their nest and an $\approx$ 1-month-old stork
were involved in the project. One of us (Z.\'A.) moved with the birds to the
Repatriation Park for 1.5 months. We fitted the three tame birds with ribbons
after they had reached the adult size. From the time they were fitted with the
ribbon, we put the GPS on them every day so they could get used to it. We
collected soaring flight data for 2 weeks before our storks joined the migrating
group. Unfortunately, the two younger storks started thermalling later than we
had expected, and very soon they joined a migrating group. Therefore, the most
of the flying tracks come from the older stork.

{\bf Collecting Flight Data from Pilots.} Paraglider, hang glider, and sailplane pilots
record their flight track with GPS during competitions, or just for themselves to
be able to overview the flight track later. These track logs are rarely available with
1-s resolution (pilots usually set their GPS devices to 3- to 5-s logging). To collect
1-s time resolution human flight track logs, we had arranged that a couple of
paraglider pilots flew in competitions with the GPS devices developed by us.
Besides, several paraglider pilots who had recorded 1-s time resolution tracks on
their own sent us their data upon our request. For some part of our research, we
did not need 1-s resolution track logs ({\it e.g.,} research related to MacCready theory).
Here, we used track logs downloaded from the Internet. The main source was the
OnlineContest (OLC) site, to which pilots from every part of the Earth upload their
track logs.

{\bf Evaluation of the Data.} The geodetic coordinates provided by the GPS were
converted into x, y, and z coordinates using the Flat Earth model. The Cubic
B-Spline method was used for fitting curves onto the points obtained with 1-s
sampling rate. The thermalling and gliding parts were separated by using
information about the curvature and the vertical velocity parameters. The
statistics was calculated from 43,000, 3,700, and 180,000 positional data
(separated in time by 1 s); 1,460, 140, and 1,430 circles in thermals; 109, 16, and
114 separate thermals for the Peregrine falcon, the White stork and the
paraglider pilots, respectively.

To determine the circling radius during thermalling, the effect of the wind had
to be eliminated. It is difficult to estimate the change of the wind force from the
ground to the top of the thermal (1,500-3,000 m) and along the horizontal route
that the birds and paragliders fly (10-100 km). Therefore, we calculated the local
wind velocity from the drifting of the thermal parts of the track log. Both
horizontal components of the velocity during thermalling were calculated. For
each component, the local maximum (minimum) places were determined. These
data were smoothed by a Gaussian filter ($\sigma =1$) and a cubic B-spline function was
fitted, providing the local maximum (minimum) velocity function. We obtained
the horizontal wind velocity component as a function of time by averaging the
local maximum and minimum velocity functions. The wind velocity component
functions were subtracted from the corresponding components of the horizontal
speed of the original track. More details about the evaluation procedures are
given in {\it SI Appendix}.

{\bf Application of the MacCready Theory.} In this theory the relation between the
horizontal and the corresponding sinking speeds (the so-called gliding polar
curve, $p(x)$, characteristic for the given gliding object) is used. It is supposed that
the climbing rate of the next thermal is known by the flyer, and no geographical
effects are taken into account. We apply the following interpretation of the
MacCready theory. The goal of the gliders is to make a given distance $L_{AB}$ (using
both thermalling and gliding and not loosing height in average) during a time as
short as possible. Thus, they intend to minimize the quantity (time) 
$L_{AB} \Bigl[ 1 / v_{xy} - {v_z / {(v_{xy} v_{climb})}} \Bigr] $, 
where $v_{xy}$, $v_z = p(v_{xy})$ are the gliding horizontal and vertical
velocities, and $v_{climb}$ denotes the climbing rate in the thermals (see Fig. \ref{fig:fig1} and 
{\it SI Appendix} Fig. 2). The optimal strategy is determined from equalling the derivative
of this expression to zero, and in this way obtaining a relationship between
the optimal $v_{xy}$ and $v_{climb}$. This leads to the expression

$${p(v_{xy})-v_{climb} \over v_{xy}}= {dp(v_{xy})\over dv_{xy}}$$

This equation is equivalent to the statement that the optimal $v_{xy}$ can be
obtained by drawing a line from the point $v_{climb}$ (along the vertical axis)
tangent to the $p(v_{xy})$ polar curve and reading the corresponding $v_{xy}$ value.

{\bf ACKNOWLEDGMENTS.} We thank our technician Mikl\'os Csisz\'er, falconer
Gy\"orgy Hank\'o, and P\'eter Zsolnay for their help in the falcon project, and
R\'obert Kiss for his contribution to work with the storks. We are also grateful
to those paraglider and hang glider pilots who shared their track logs with us.
We especially thank the Hortob\'agy National Park for providing the place and
giving permission for our research. This research was partially supported by
Orsz\'agos Tudom\'anyos Kutat\'asi Alapprogramok (OTKA) 049674 and the European
Union FP6 STARFLAG projects.


\begin{thebibliography}{99}

\bibitem{bib1}
Pennycuick CJ (1971) Gliding flight of the white-backed vulture Gyps africanus. {\it J Exp
Biol} 55:13-38.

\bibitem{bib2}
Pennycuick CJ (1983) Thermal soaring compared in three dissimilar tropical bird
species, Fregata magnificens, Pelecanus occidentalis and Coragyps atratus.  {\it J Exp Biol}
102:307-325.

\bibitem{bib3}
Shamoun-Baranes J, Leshem Y, Yom-Tov Y, Liechti O (2003) Differential use of thermal
convection by soaring birds over central Israel. {\it Condor} 105:208-218.

\bibitem{bib4}
Shamoun-Baranes J, van Gasteren JR, van Belle J, Bouten W, Buurma LS (2005) Flight
altitudes of birds.  {\it Bull Amer Meteorol Soc} 86:18-19.

\bibitem{bib5}
Shannon HD, Young GS, Yates M, Fuller MR, Seegar W (2003) American white pelican
soaring flight times and altitudes relative to changes in thermal depth and intensity.
{\it Condor} 104:679-683.

\bibitem{bib6}
Weimerskirch H, Chastel O, Barbraud C, Tostain O (2003) Frigatebirds ride high on
thermals. {\it Nature} 421:333-334.

\bibitem{bib7}
MacCready PB (1958) Optimum airspeed selector. {\it Soaring} 10-11.

\bibitem{bib8}
Spaar R, Bruderer B (1997) Migration by flapping or soaring: Flight strategies of Marsh,
Montagu's and Pallid harriers in Southern Israel. {\it Condor} 99:458-469.

\bibitem{bib9}
Tucker VA (2001) Gliding birds: The effect of variable wing span. {\it J Exp Biol} 133:33-58.

\bibitem{bib10}
Lipp H-P, et al. (2004) Pigeon homing along highways and exits. {\it Curr Biol} 14:1239-1249.

\bibitem{bib11}
Biro D, Meade J, Guilford T (2004) Familiar route loyalty implies visual pilotage in the
homing pigeon. {\it Proc Natl Acad Sci USA} 101:17440-17443.

\bibitem{bib12}
Ganusevich SA, et al. (2004) Autumn migration and wintering areas of Peregrine Falcons
(Falco peregrinus) nesting on the Kola Peninsula, northern Russia. {\it Ibis} 146:291-297.

\bibitem{bib13}
Lentink D, et al. (2007) How swifts control their glide performance with morphing
wings. {\it Nature} 446:1082-1085.

\bibitem{bib14}
Chatterjee S, Templin RJ, Campbell KE, Jr (2007) The aerodynamics of Argentavis, the
world's largest flying bird from the Miocene of Argentina. {\it Proc Natl Acad Sci USA}
104:12398-12403.

\bibitem{bib15}
Cochrane JH (1999) MacCready theory with uncertain lift and limited altitude. {\it Technical
Soaring} 23:88-96.

\end{thebibliography}
\end{document}